\documentclass[a4paper,11pt]{article}
\pdfoutput=1
\usepackage{jcappub} 
\usepackage[T1]{fontenc}

\usepackage{amsmath,amssymb,amsfonts,bm,mathtools}
\allowdisplaybreaks
\usepackage{graphicx}
\usepackage{booktabs}
\usepackage{tabularx}
\usepackage{float}
\usepackage{subcaption}
\usepackage{pifont}
\usepackage{enumerate}
\usepackage{microtype}
\usepackage[table,xcdraw]{xcolor}
\usepackage{hyperref}
\usepackage{natbib}
\usepackage{tikz}
\usepackage{tablefootnote}
\usepackage{afterpage}

\newcommand{\orcid}[1]{\href{https://orcid.org/#1}{\,\includegraphics[width=8px]{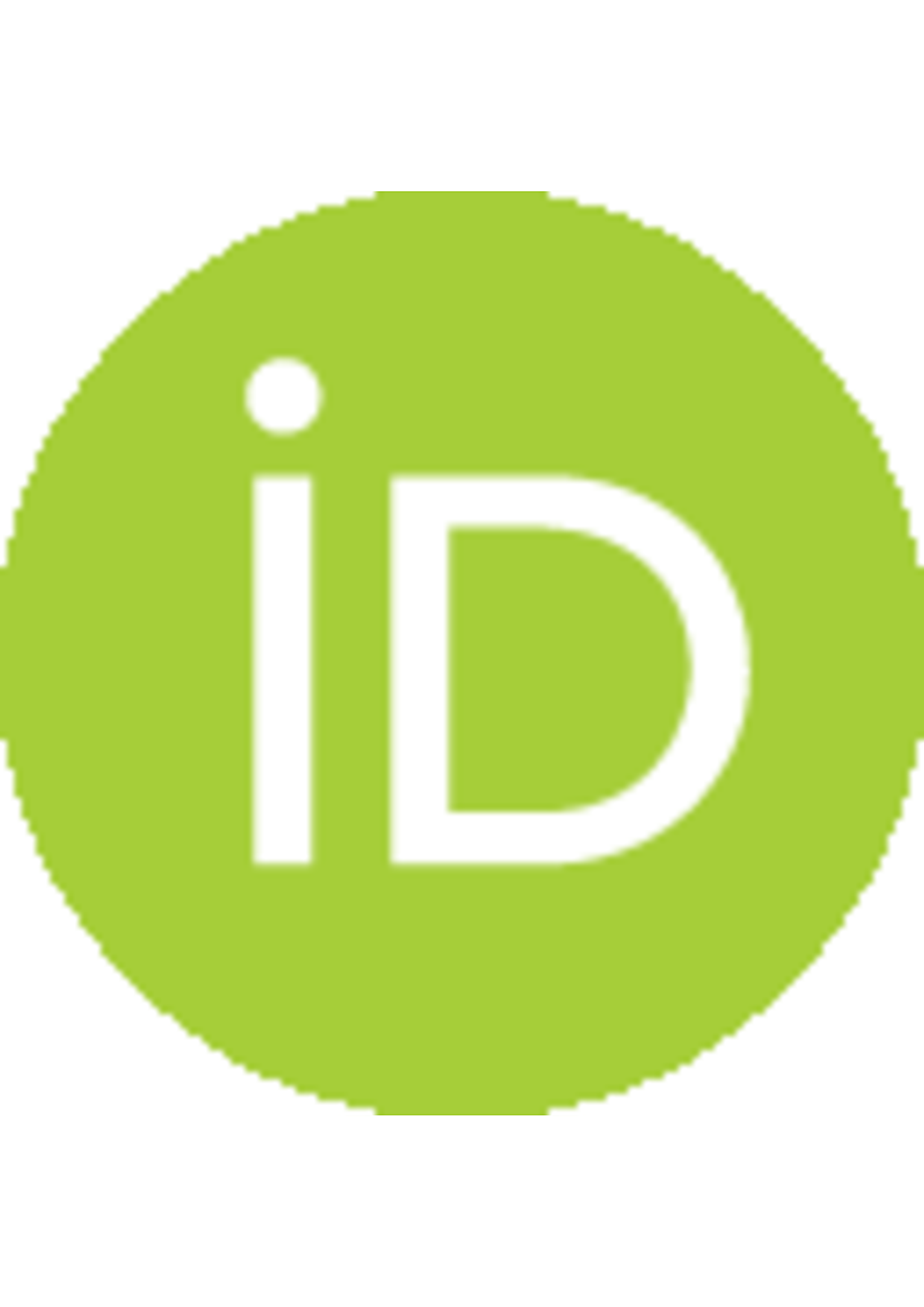}}}

\usepackage{soul}
\setstcolor{blue}
\usepackage{cancel}

\usepackage{etoolbox}

\definecolor{lightgray}{gray}{0.9}

\def\be{\begin{equation}}
\def\ee{\end{equation}}
\def\bea{\begin{eqnarray}}
\def\eea{\end{eqnarray}}
\def\dd{{\rm d}}

\DeclareMathAlphabet{\mathpzc}{OT1}{pzc}{m}{it}

\allowdisplaybreaks

\title{\textbf{Impact of lensing magnification on the power spectrum turnover}}

\author[a]{Yolanda Dube,\!\orcid{0000-0002-8882-5474}}
\author[a,b]{Roy Maartens,\!\orcid{0000-0001-9050-5894}}
\author[a]{Bikash R. Dinda,\!\orcid{0000-0001-5432-667X}}
\author[a]{\\ S\^ecloka L. Guedezounme,\!\!\! \orcid{0009-0000-9200-8584}}
\author[a]{Sheean Jolicoeur}

\affiliation[a]{Department of Physics \& Astronomy, University of the
Western Cape, Cape Town 7535, South Africa}
\affiliation[b]{National Institute for Theoretical \& Computational
Sciences, Cape Town 7535, South Africa}

\emailAdd{yolymatics007@gmail.com}
\emailAdd{rmaartens@uwc.ac.za}
\emailAdd{bikashrdinda@gmail.com}
\emailAdd{seclokaguedezounme@gmail.com}
\emailAdd{jolicoeursheean@gmail.com}

\abstract{
The turnover scale $k_0$ of the matter power spectrum -- and consequently of the standard galaxy power spectrum monopole -- encodes a fundamental signature of matter--radiation equality and constitutes an important standard ruler independent of baryon acoustic oscillations. In principle, we can detect the turnover at multiple redshifts and amplify the signal by stacking redshift bins. However, in spectroscopic surveys reaching high redshifts, such as the Euclid H$\alpha$ survey and the proposed MegaMapper Lyman-break galaxy survey, the monopole of the observed galaxy power spectrum receives a scale-dependent correction from lensing magnification. This can modify the signal shape and shift the turnover scale, undermining its use as a standard ruler. Using mock surveys similar to Euclid and MegaMapper, we forecast this shift and the consequent bias in the turnover scale that is recovered from the mock data. The shift in the turnover scale grows with redshift. In units of the precision of the stacked no-lensing analysis, the shift reaches $\sim 0.3$ (Euclid-like) and $\sim 3.5$ (MegaMapper-like). For a MegaMapper-like survey, the shift in a single redshift bin exceeds this precision for $z\gtrsim 2.9$, and the turnover vanishes for $z\gtrsim 3.7$. Both redshifts depend on the adopted magnification bias: a 20\% change in $s(z)$ moves them by $\Delta z\approx0.3$ or more. Our results show that the lensing correction to the monopole should be included and carefully modelled when surveys are used to measure the cosmological turnover at high redshifts.
}

\begin{document}

\maketitle
\tableofcontents
\flushbottom


\notoc
\section{Introduction}
\label{sec:intro}

State-of-the-art spectroscopic surveys are mapping the three-dimensional distribution of galaxies over unprecedented volumes, opening a window to the largest scales in the Universe. Among the features accessible on these scales is the turnover of the matter power spectrum at comoving wavenumber $k_0 \approx 0.0163\,
h/\mathrm{Mpc}$, which marks the transition between modes that entered the Hubble horizon during radiation and matter domination. Because $k_0$ is set by the physics of matter--radiation equality, it serves in principle as a standard ruler that is complementary to baryon acoustic oscillations (BAO). A precise measurement of $k_0$ can constrain the physical matter density $\Omega_{m0} h^2$ and provide an independent route to the Hubble constant $H_0$~\cite{prada2011measuring,poole2013wigglez,Pryer:2021cut,Cunnington:2022ryj,Bahr-Kalus:2023ebd, Alonso:2024emk,DESI:2025euz}.

In our recent paper~\cite{Dube:2025hqf}, we performed MCMC forecasts for high-redshift spectroscopic surveys similar to the Euclid H$\alpha$ survey and the proposed MegaMapper survey. The motivation is that the greater cosmic volume at high $z$ will enhance the precision on turnover measurement. However, in addition to precision, there is also the need for accuracy. In particular, we need to check that the theory includes all relevant effects.
In common with most previous work, in~\cite{Dube:2025hqf} we modeled the  monopole of the  galaxy power spectrum using the standard (Kaiser) approximation~\cite{Kaiser:1987qv} in the observed galaxy density contrast:
\begin{align} \label{obss}
    \delta_{{\rm obs}}=
    \delta_{\rm S} =
    b\,\delta_m-\frac{1}{\cal H}\hat{\bm n}\cdot\bm{\nabla}\big( \hat{\bm n}\cdot{\bm v}\big)\,,
\end{align}
where $b$ is the linear clustering bias,  ${\cal H}$ is the conformal Hubble rate, $\hat{\bm n}$ is the line of sight, $\bm v$ is the peculiar velocity and $\delta_m$ is the matter  density contrast. The dominant correction to this approximation for wide and deep surveys is the lensing magnification contribution to the observed power spectrum~\cite{Bonvin:2011bg,Challinor:2011bk,Jeong:2011as} (see also~\cite{Camera:2014sba,Alonso:2015uua,Jelic-Cizmek:2020pkh,Grimm:2020ays,Castorina:2021xzs,Viljoen:2021ocx,Lepori:2021lck,Noorikuhani:2022bwc,Foglieni:2023xca,Euclid:2023qyw,Guedezounme:2024pbj, Addis:2025rre}). Weak gravitational lensing by foreground structure modifies the standard observed galaxy number density contrast via the lensing convergence   $\kappa$:
\begin{align} \label{dobs}
    \delta_{{\rm obs}}=
    \delta_{\rm S} + \delta_{\rm L}=\delta_{\rm S}+ \big(5s-2 \big) \kappa\,.
\end{align}
Here $s$ is the magnification bias, determined by the slope of the luminosity function at the survey magnitude limit:
\begin{align}\label{sdef}
s=\frac{\partial\log_{10}N(<m)}{\partial m}\bigg|_{m_c}\,,
\end{align}
where $N(<m)$ is the number density of sources brighter than apparent magnitude $m$, and $m_c$ is the magnitude limit of the survey. An equivalent convention, used in~\cite{Addis:2024bss,Addis:2025rre}, is $\mathcal{Q}=-\partial\ln N(>L)/\partial\ln L$ at the luminosity threshold, with $\mathcal{Q}=\tfrac52 s$, so that $5s-2=2(\mathcal{Q}-1)$. The lensing convergence  is
\begin{align}
  \kappa(\hat{\bm{n}}, z)
  = -\frac{1}{2}
     \int_0^{r(z)} \dd r'\,
     \frac{[r(z)-r']\,r'}{r(z)}
     \nabla_\perp^2\Phi(r'\hat{\bm{n}}, z') \,,
  \label{eq:kappa}
\end{align}
where $r$ is the comoving radial distance, $\nabla_\perp^2$ is the transverse Laplacian and $\Phi$ is the gravitational potential.

It is known that ignoring the lensing effect can bias the measurement of local primordial
non-Gaussianity (e.g. \cite{Camera:2014sba,Alonso:2015uua,Jelic-Cizmek:2020pkh,Viljoen:2021ocx,Addis:2025rre}). This underscores the importance of lensing corrections for analysis on ultra-large scales. In high-redshift surveys such as the Euclid H$\alpha$ survey ($0.9\leq z\leq 1.8$) and the proposed MegaMapper Lyman-break galaxy survey ($2.1\leq z\leq 5$), the lensing convergence integral accumulates contributions over very long lines of sight, and the resulting correction to the power spectrum monopole can become non-negligible on the scales relevant to the turnover.

In \cite{LoVerde:2007ke,Hui:2007tm}, lensing effects on the 2-point correlation function, the angular power spectrum, and the Fourier power spectrum of galaxies were investigated. In particular, they noted a shift in the matter-radiation equality scale due to lensing. Here, we investigate quantitatively for the first time the impact of lensing magnification on the detection and measurement of the matter power spectrum turnover $k_0$. We use Euclid-like and MegaMapper-like surveys as examples. Our main findings are as follows.

\begin{itemize}
    \item
The intrinsic turnover scale $k_0$ in the matter power spectrum is shifted by lensing effects in the observed galaxy monopole:
\begin{align} \label{tko}
    k_0~\to~ \tilde{k}_0(z)\,. 
\end{align}
The shift typically grows with redshift.
\item This shift therefore leads to a biased measurement of the intrinsic (matter) turnover scale. In the case of a Euclid-like survey, the turnover is still present, but biased, at the maximum redshift. By contrast, the turnover in the case of a MegaMapper-like survey disappears for all redshifts above $z\approx 3.7$, where the precise $z$-value depends on the model of magnification bias $s(z)$, on the $k$-range used to detect the turnover, and on how a detectable turnover is defined. For our fiducial $s(z)$ we find $3.5\lesssim z\lesssim 3.8$.

\item As a result, high redshift spectroscopic surveys may not deliver a `clean' probe of the cosmological ruler encoded in the turnover scale. To estimate and, if necessary, remove the lensing effect on $k_0$, the magnification bias in such surveys must be carefully modeled.

\item
We express the shift in a redshift bin, $k_0-\tilde k_{0}(z_i)$, in units of the precision $\sigma(k_0)$ that is obtained by stacking all bins when lensing is neglected. In the Euclid-like case this ratio remains below 1, with a maximum of $\sim 0.3$. In the MegaMapper-like case it exceeds 1 for $z\gtrsim 2.9$, and reaches $\sim 3.5$ before the turnover disappears. Measurements above $z\approx2.9$ therefore do not provide a reliable estimate of the intrinsic turnover, unless the lensing effect can be accurately modeled and removed. If lensing is neglected and all MegaMapper-like bins up to $z=3.6$ are stacked, the recovered $k_0$ is biased by $-0.93\sigma$.

\end{itemize}

We adopt the Planck 2018 flat $\Lambda$CDM parameters. The matter power spectrum is computed with \texttt{CLASS}~\cite{Blas:2011rf},
and MCMC sampling is performed with \texttt{emcee}~\cite{Foreman-Mackey:2013rvk}.
Power spectra {and associated covariances} including lensing are computed using \texttt{CosmoWAP}~\cite{Addis:2024bss,
Addis:2025rre}.

\section{{Lensing correction to the observed galaxy power spectrum monopole}}
\label{sec:theory}

The linear matter power spectrum is
\begin{equation}
    P_m(k,z) = D^2(z)\,P_m(k,0) \,,
    \label{eq:matter_pm}
\end{equation}
where $D$ is the growing mode of the linear growth factor, normalised to $D(0)=1$, $P_m(k,0) = A\, k^{n_s} T^2(k)$, where $A$ is determined by the amplitude of the primordial power spectrum, $n_s$ is its spectral tilt and $T(k)$ is the matter transfer function, encoding the effects of the matter--radiation
transition. Modes inside the comoving Hubble horizon at equality ($k > k_{\rm eq}$) are suppressed relative to super-horizon modes ($k < k_{\rm eq}$). On super-horizon scales, $T(k) \sim 1$ and $P_m \sim k^{n_s}$, while on subhorizon scales, $T(k) \sim k^{-2}$ and
$P_m \sim k^{n_s - 4}$. The transition between these regimes imprints the peak, or turnover,
in $P_m$, at a scale $k_0$, which is independent of redshift.

In the standard  approximation, \autoref{obss}, the  galaxy power spectrum monopole
in redshift space is
\begin{equation}
P_{\rm S}^{(0)}(k, z)
= \left[ b^2(z) + \frac{2}{3}\,b(z)\,f(z) + \frac{1}{5}\, f^2(z)\right]
  P_m(k, z) \,,
\label{eq:monopole}
\end{equation}
where $f = - \dd \ln D/\dd \ln(1+z)$
is the linear growth rate. The standard monopole \eqref{eq:monopole} captures the dominant contribution to the power spectrum in most cases of interest. However, in a flux-limited survey probing high redshifts, gravitational lensing introduces a correction to this.

In Fourier space, the lensing contribution given by \autoref{dobs} and \autoref{eq:kappa} can be expressed through a kernel
$\mathcal{K}_{\rm L}$:
\begin{align}
 \delta_{\rm L}(\bm{x})
  = \int_0^{x} \dd r' \int \frac{{\rm d}^3\bm{q}}{(2\pi)^3}\,\,
    \mathrm{e}^{{\rm i}\,\bm{q}\cdot\bm{r}'}\,
    \mathcal{K}_{\rm L}(\bm{q}, \bm{x}(z), r')\,\delta_0(\bm{q}) \,,
  \label{eq:DeltaL_fourier}
\end{align}
where $\delta_0$ is the linear matter overdensity at $z=0$, $x = r(z)$ is the comoving distance to the source, with $\bm x=x\, \hat{\bm n}$, and $r'$ runs along the line of sight from observer to source. The lensing kernel is~\cite{Addis:2025rre}
\begin{align}
  \mathcal{K}_{\rm L}(\bm{q}, {\bm x}(z), r')
  = {\frac32\,[5s(z) - 2]\,D(r')}\,\Omega_m(r')\,\mathcal{H}^2(r')\,
    \frac{(x(z) - r')\,r'}{x(z)}
    \left[1 - (\hat{\bm{q}}\cdot\hat{\bm{x}}(z))^2
      +2\,{\rm i} \, \frac{\hat{\bm{q}}\cdot\hat{\bm{x}}(z)}{r'q}
    \right] .
  \label{eq:kernel_L}
\end{align}
The real part of the angular factor, $1 - (\hat{\bm{q}}\cdot\hat{\bm{x}})^2$, arises from the transverse Laplacian in the convergence
integral, \autoref{eq:kappa}, while the imaginary part is a wide-angle correction. The magnification bias is a property of the source sample, and it is evaluated at the source redshift $z$, while $D$, $\Omega_m$ and $\mathcal{H}$ are evaluated along the line of sight. Note that this wide-angle correction is, in general, dependent on the line-of-sight choice (endpoint, midpoint, etc) -- but this dependence is removed by the averaging that produces the monopole of the power spectrum~\cite{Addis:2025rre}.

The standard monopole, \autoref{eq:monopole}, is corrected by the cross-power between the standard term and the lensing convergence, and by the lensing auto-power:
\begin{align}\label{eq:Pobs}
P^{(0)}_{\rm obs}=P^{(0)}_{\rm S}+P^{(0)}_{\rm S\times L}+P^{(0)}_{\rm L\times L}\,.
\end{align}
The cross-power $P^{(0)}_{\rm S\times L}$ (which includes both ${\rm S}\times{\rm L}$ and ${\rm L}\times{\rm S}$) involves a single line-of-sight integral of the matter power spectrum weighted by the lensing kernel, \autoref{eq:kernel_L}, while the auto-power $P^{(0)}_{\rm L\times L}$ involves a double line-of-sight integral. Both are evaluated after integrating over the angle {$\mu=\hat{\bm{q}}\cdot\hat{\bm{x}}$}
and averaging over the survey volume. The numerical computation of these lensing contributions is performed with \texttt{CosmoWAP}~\cite{Addis:2025rre}. In the turnover region, $P^{(0)}_{\rm L\times L}$ dominates the correction: it is 10--30 times larger than $P^{(0)}_{\rm S\times L}$, and grows towards small $k$. For example, at $k=0.006\,h\,{\rm Mpc}^{-1}$ we find $P^{(0)}_{\rm L\times L}/P^{(0)}_{\rm S}\approx0.13$ for the Euclid-like survey at $z=1.69$, and $\approx0.47$ for the MegaMapper-like survey at $z=3.55$.

The measurement covariance is similarly modified when lensing is included. For the Gaussian covariance of the monopole we use
\begin{align}\label{eq:cov}
{\rm Cov}\big[P^{(0)}_{\rm obs}\big](k)&=\frac{1}{N_k}\int_{-1}^{1}{\rm d}\mu\,\Big[P_{\rm S}(k,\mu)+P_{\rm S\times L}(k,\mu)+P_{\rm L\times L}(k,\mu)+\frac{1}{n_g}\Big]^2,
\notag \\
\text{where}~~ N_k&=\frac{V k^2\Delta k}{2\pi^2}\,.
\end{align}
Here $n_g$ is the comoving number density, $V$ is the volume of the redshift bin and $\Delta k$ is the $k$-bin width. The lensing terms enter through their cross-products with $P_{\rm S}$ and the shot noise, and through their squares. The $\mu$-dependent spectra are computed with \texttt{CosmoWAP}~\cite{Addis:2025rre}, and without lensing \autoref{eq:cov} reduces to the standard \texttt{CosmoWAP} monopole covariance.

\begin{figure}[!htbp]
  \centering
  \includegraphics[width=0.97\linewidth]
    {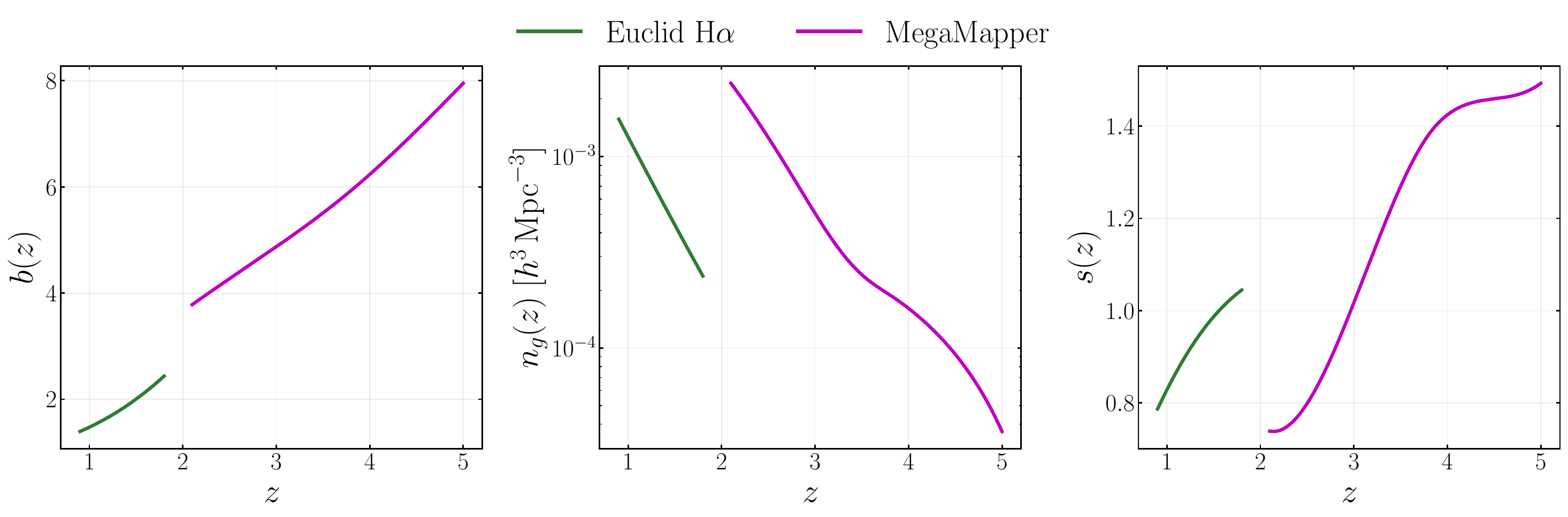}
  \caption{
    Linear clustering bias $b$,  comoving number density $n_g$, and magnification bias $s$, for the Euclid-like (green) and MegaMapper-like (magenta) surveys, as functions of redshift. (Note that a typo in the MegaMapper-like clustering bias in \cite{Dube:2025hqf,Rossiter:2024tvi} is corrected in \cite{Addis:2025rre}.)
  }
  \label{fig:survey_params}
\end{figure}

The specifications for the Euclid-like and  MegaMapper-like surveys are taken from
\cite{Euclid:2019clj,Schlegel:2022vrv,Rossiter:2024tvi,Addis:2025rre}. In particular, to compute the theoretical power spectra and covariances, we need the clustering bias, magnification bias, and comoving number density as functions of redshift. Simplified models for these quantities are shown in \autoref{fig:survey_params}. The magnification bias is computed with \texttt{CosmoWAP} from the luminosity function at the survey threshold. For the Euclid-like survey we use the H$\alpha$ luminosity function Model~3 of~\cite{Pozzetti:2016cch}, with flux limit $2\times10^{-16}\,{\rm erg\,s^{-1}cm^{-2}}$. For the MegaMapper-like survey we use the Lyman-break galaxy UV luminosity functions of~\cite{Wilson:2019brt}, fitted at $z=2.0$, 3.0, 3.8, 4.9 and 5.9 and interpolated in redshift, with magnitude limit $m_c=24.5$.

\newpage
\section{Bias on the turnover from neglecting lensing}
\label{sec:results}

\begin{figure}[!htbp]
  \centering
  \includegraphics[width=\linewidth]%
    {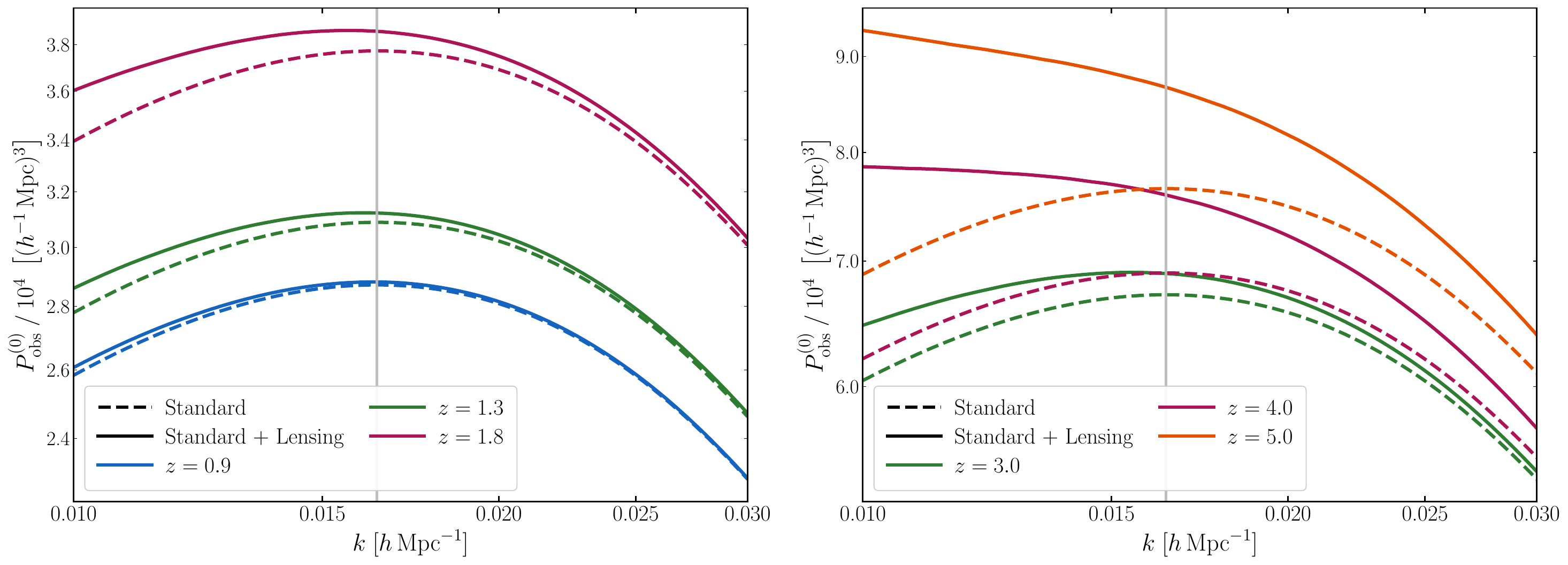}
  \caption{
    The fiducial galaxy power spectrum monopoles in the turnover region for Euclid-like (left) and  MegaMapper-like (right) surveys at various redshifts. Dashed lines are the standard model, and solid lines are the lensing-corrected model. The fiducial turnover scale $k_0$  is shown by the grey lines.}
  \label{fig:monopoles_full_range}
\end{figure}

To begin with, we illustrate in \autoref{fig:monopoles_full_range} the shift in the turnover position due to lensing, by computing the monopoles of the power spectra without lensing (dashed curves) and with the lensing correction (solid curves), for some example redshifts.  It is apparent that for these surveys, the lensing contribution leads to a growth of power, especially on very large scales, $k\lesssim k_{\rm eq}$. The effect increases with redshift. For the Euclid-like case, there is clearly a bias in turnover measurement: the effect is not negligible, but not large.  The MegaMapper-like case shows a significant shift in the turnover, with no turnover at the 2 highest redshifts.

\begin{figure}[!htbp]
  \centering
  \includegraphics[width=0.7\linewidth]
    {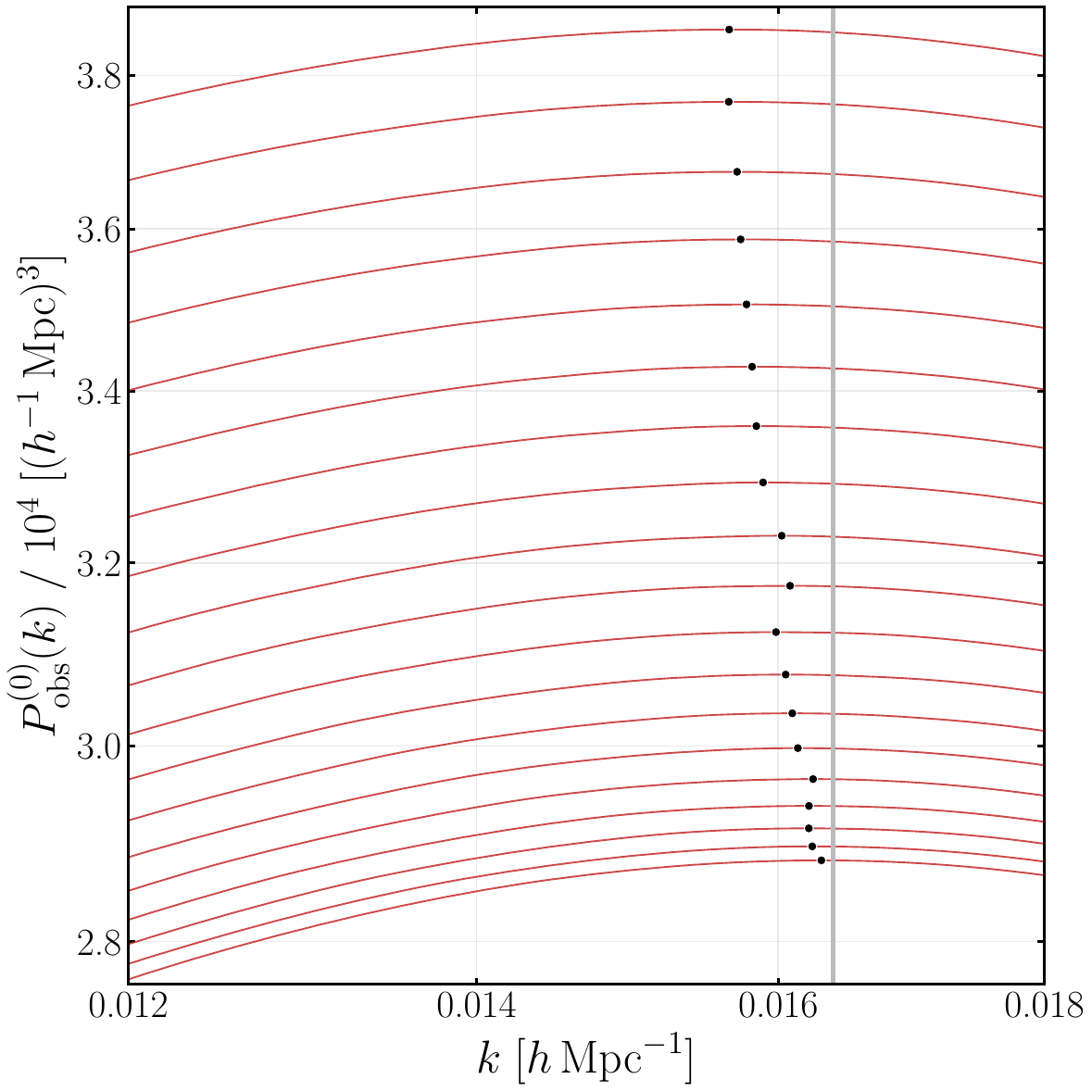}
  \caption{
    Lensing-induced shift of the turnover for a
    Euclid-like survey. Dots indicate the turnover at redshifts from the minimum to the maximum $z$ in steps of 0.1. The fiducial turnover scale $k_0$ is shown by the grey line.
    }
  \label{fig3}
\end{figure}

\begin{figure}[!htbp]
  \centering
  \includegraphics[width=0.49\linewidth]%
    {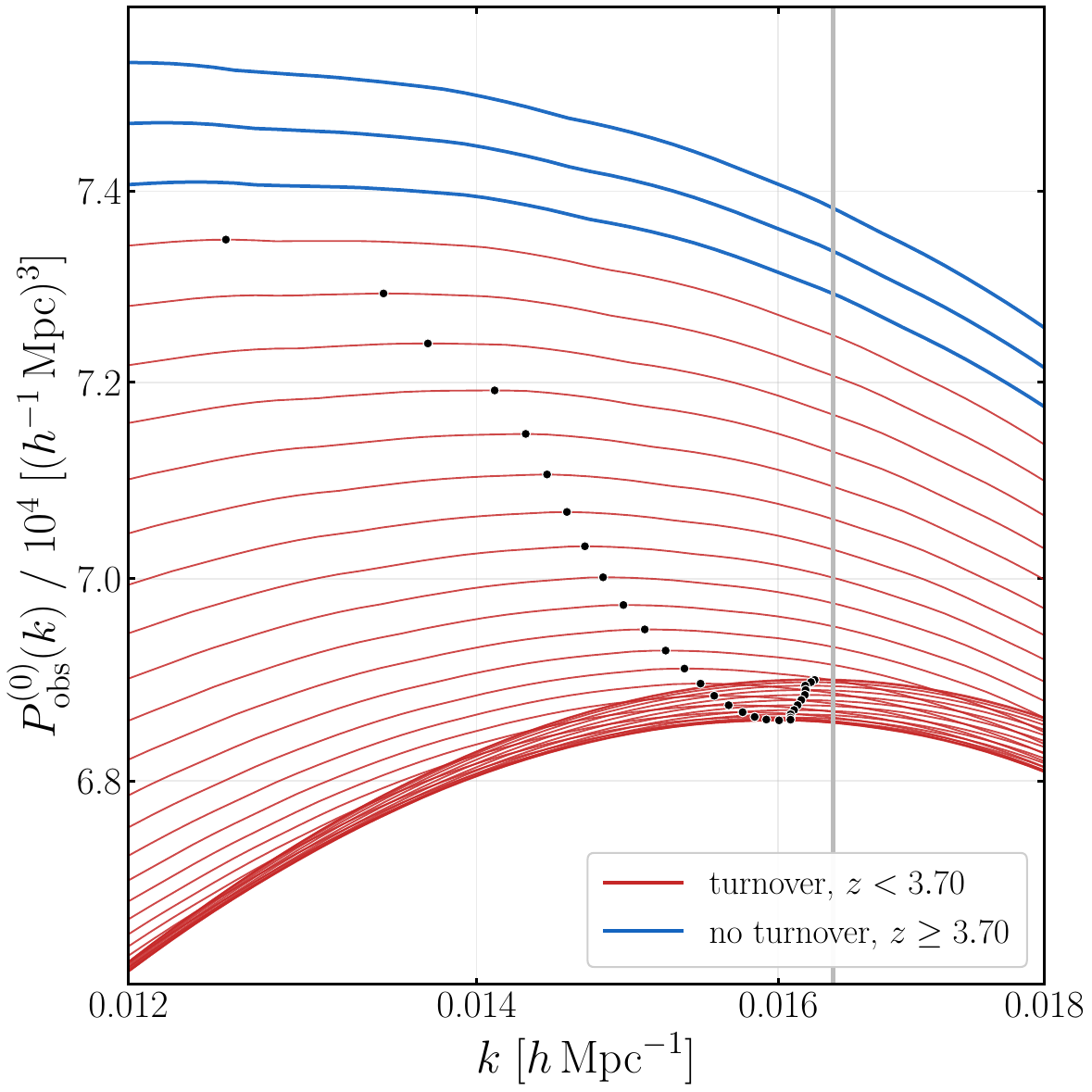}%
  \hfill
  \includegraphics[width=0.49\linewidth]%
    {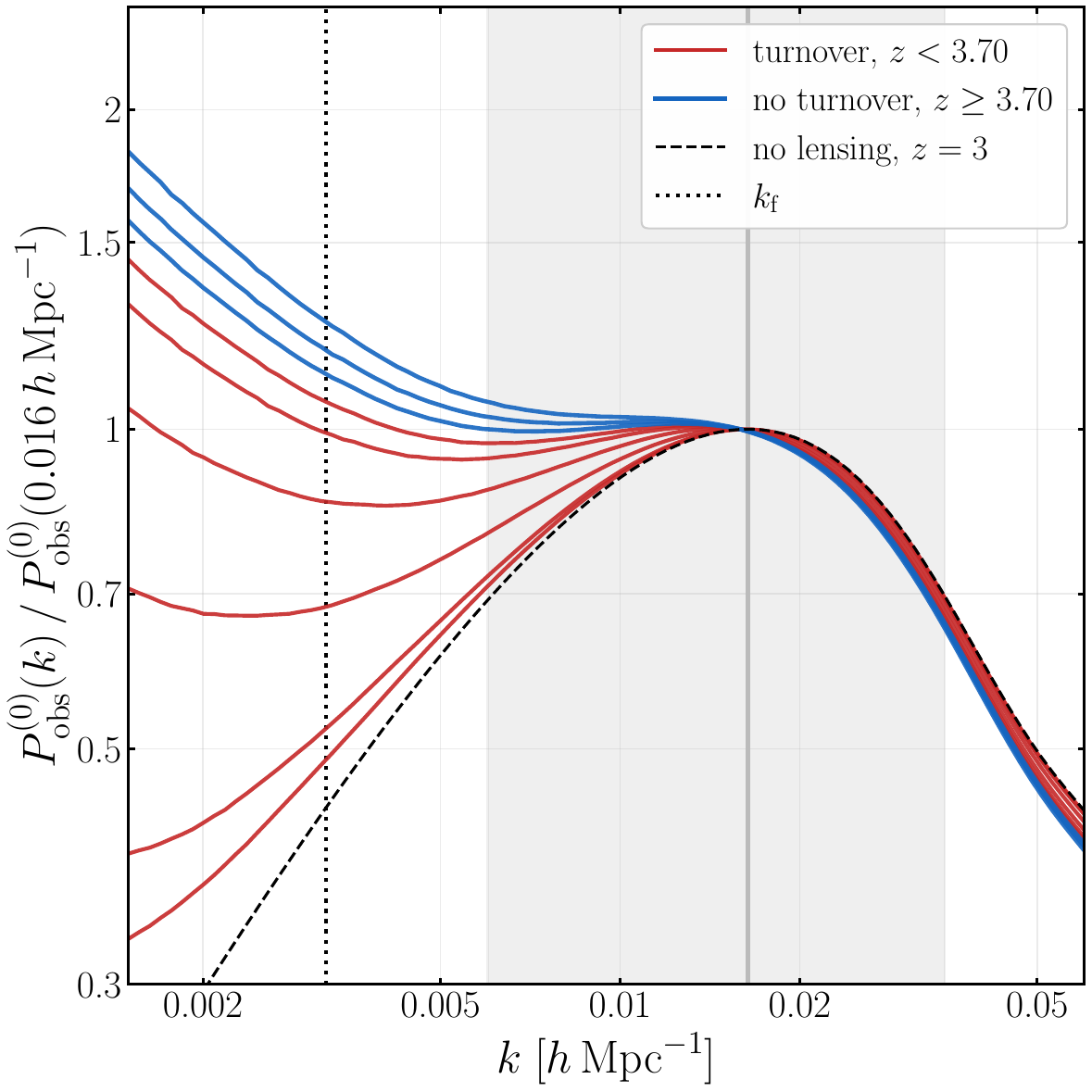}%
  \caption{
    Lensing-induced shift of the turnover for a MegaMapper-like survey. {\em Left:} Dots indicate the turnover at redshifts from the minimum to the maximum $z$ in steps of 0.05.  Red curves show the redshifts at which the turnover is still present. The dot trail terminates at  $z =3.65$. Blue curves are the first three redshifts past the disappearance threshold $z = 3.70$. The fiducial turnover scale $k_0$ is shown by the grey line. {\em Right:} The monopole over the full range $0.0015\le k\le0.06\,h/$Mpc, normalised at $k=0.016\,h/$Mpc, for $z=2.1$, 2.5, 3.0, 3.3, 3.5, 3.6 (red) and $z=3.7$, 3.8, 3.95 (blue). The shaded band is the fitting window, the dotted line is the fundamental mode $k_{\rm f}$ of a redshift bin, and the dashed curve is the standard monopole at $z=3$.
    }
  \label{fig4}
\end{figure}

A more systematic illustration is given in \autoref{fig3} and \autoref{fig4}. These show the shifted turnover $\tilde k_0(z)$ (marked by black dots) explicitly at a series of redshifts. For MegaMapper-like, the lensing contribution is so strong that the turnover disappears for $z\geq 3.70$. Here we define the turnover as a maximum of $P^{(0)}_{\rm obs}$ inside the fitting window $0.006\le k\le0.035\,h/$Mpc, which rises and falls by at least 2\% towards the edges of the window. The right panel of \autoref{fig4} shows that the turnover does not move out of the window. Instead, $P^{(0)}_{\rm L\times L}$ grows steeply towards small $k$, so that $P^{(0)}_{\rm obs}$ turns up again below the turnover and the peak is washed out. We note that this redshift value is sensitive to the modelling of magnification bias $s(z)$, as we show below.

To estimate the bias on turnover measurement that is induced by neglecting lensing, we need to express $k_0- \tilde k_0(z)$ in units of the error $\sigma(k_0)$. First, we need to recover $k_0$ and its error $\sigma(k_0)$, i.e., in the {\bf no-lensing} case. For this case, we
use the model-independent approach of~\cite{poole2013wigglez,Cunnington:2022ryj,
Bahr-Kalus:2023ebd,Alonso:2024emk}, also implemented in \cite{Dube:2025hqf}, which fits an asymmetric parabola model to the standard monopole in the neighbourhood of the  turnover:
\begin{align}
  \log P_{\rm S}^{\rm model}(k,{z_i})
  = \log P_{\rm S}^{(0)}(k_0,z_i)
  \times\begin{cases}
    \big(1 - {\alpha} x^2\big) & k \leq k_0 \\[6pt]
    \big(1 - {\beta} x^2\big) & k >    k_0
    \end{cases} \,,
  \quad
  x = \frac{\log(k/k_0)}{\log\big(k_0/1\,h\mathrm{Mpc}^{-1}\big)} \, ,
  \label{eq:parabola}
\end{align}
where $P_{\rm S}^{(0)}(k_0,z_i)$ is the peak value of the standard power spectrum monopole at $z=z_i$, given by \autoref{eq:monopole}. The free parameters in each bin, $\theta_{ai}=(k_{0,i},\alpha_i,\beta_i,P_{0,i})$, are the turnover scale, the curvature slopes and the peak amplitude $P_{0,i}=P_{\rm S}^{(0)}(k_0,z_i)$. We fit each redshift bin separately. Since the turnover scale is redshift-independent when lensing is not included, we then combine the per-bin turnover scales into a single $k_0$ by inverse-variance weighting:
\begin{align}\label{eq:stack}
k_0=\sigma^2(k_0)\sum_i\frac{k_{0,i}}{\sigma^2(k_{0,i})}\,,\qquad \sigma^{-2}(k_0)=\sum_i\sigma^{-2}(k_{0,i})\,,
\end{align}
where $k_{0,i}$ is the maximum of the marginal posterior in bin $i$, and $\sigma(k_{0,i})$ is half the width of its 68\% interval.

When {\bf lensing} is included, the turnover is redshift-dependent, $k_0\to \tilde k_0(z)$, and we use the standard + lensing  monopole, so that the parameters change in each redshift bin:
\begin{align}
\log P_{\rm S+L}^{\rm model}(k,{z_i})
  = \log P_{\rm S+L}^{(0)}(\tilde k_{0,i},z_i)
  \times\begin{cases}
    \big(1 - {\alpha_i} x_i^2\big) & k \leq \tilde k_{0,i} \\[6pt]
    \big(1 - {\beta_i} x_i^2\big) & k >    \tilde k_{0,i}
    \end{cases} \,,
  \quad~~
  x_i = \frac{\log\big(k/{\tilde k_{0,i}}\big)}{\log\big({\tilde k_{0,i}}/1\,h\,\mathrm{Mpc}^{-1}\big)} \!.
  \label{eq:parabola2}
\end{align}
As in the standard case, the parameters are fitted separately in each redshift bin, $\theta_{ai}=\big(\tilde k_{0,i},\alpha_i,\beta_i,\tilde P_{0,i}\big)$, but now the turnover scale itself depends on redshift.
Here $ P_{\rm S+L}^{(0)}(\tilde k_{0,i},z_i)$ is the peak value at $k=\tilde{k}_{0,i}$ in the $i$ redshift bin of the lensing corrected power spectrum monopole.
We adopt  Gaussian likelihoods:
\begin{align}
\label{eq:chi}
  \bm{\chi}^2_{\rm S}(z_i)
  &= \sum_{j}
    \frac{\left[P^{(0)}_{\rm obs}(k_j, z_i)
    - P_{\rm S}^{\rm model}(k_j;\,\theta_{ai})\right]^2}
    {\mathrm{Var}[P^{(0)}_{\rm obs}](k_j, z_i)} \, , \\
    \label{eq:chi2}
      \bm{\chi}^2_{\rm S+L}(z_i)
  &= {\sum_{j}
    \frac{\left[P^{(0)}_{\rm obs}(k_j, z_i)
    - P_{\rm S+L}^{\rm model}(k_j;\,{\theta_{ai}})\right]^2}
    {\mathrm{Var}[P^{(0)}_{\rm obs}](k_j, z_i)} \, , }
\end{align}
where `Var' corresponds to the variance which is the auto covariance computed using \autoref{eq:cov}, with the lensing terms included only for the lensed data.

In each redshift bin, the comoving survey volume is
\begin{align}
  V_{i} = \frac{4\pi}{3}\,f_{\rm sky}
    \Big[r^3(z_{i}+\Delta z/2) - r^3(z_{i}-\Delta z/2)\Big] \,,
\end{align}
where $f_{\rm sky}$ is the sky fraction. The volume defines the fundamental mode in the bin,
$k_{\rm f} = 2\pi V_{i}^{-1/3}$ and the $k$-bin width is set to ${\Delta k=  2k_{\rm f}}$. In order to avoid contamination from the BAO and remain within the linear regime, we chose the shortest mode as $k_{\rm max}= 0.035 h/$\,Mpc. For the minimum, we use $k_{\rm min}= 0.006h$/Mpc. For the MegaMapper-like bins, $k_{\rm f}\approx0.0032\,h/$Mpc, so that $k_{\rm min}\approx1.9\,k_{\rm f}$ and each bin contains 6 $k$-values. The dependence of our results on $k_{\rm min}$, $k_{\rm max}$ and $\Delta k$ is discussed below. The data vector in each bin consists of the theoretical power spectrum monopole -- without lensing in the standard case, and with lensing in the lensed case.

In order to assess the effect of lensing on the measured turnover, we perform MCMC sampling as follows.
\begin{itemize}
    \item
Without lensing, we find the recovered turnover $k_0$ and the precision $\sigma(k_0)$, by stacking all redshift bins with \autoref{eq:stack}. Since each redshift bin has a certain constraining power on $k_0$ and all the redshift bins constrain the same parameter $k_0$, the stacking of all redshift bins puts the best possible constraint on the turnover scale $k_0$.
    \item
With lensing, we find the turnover in each redshift bin, $\tilde k_0(z_i)\equiv\tilde k_{0,i}$, and the lensing-induced shift $k_0-\tilde k_{0,i}$. We express this shift in units of the precision of the stacked no-lensing analysis:
\begin{align}\label{eq:D}
\mathcal{D}_i\equiv\frac{k_0-\tilde k_{0,i}}{\sigma(k_0)}\,.
\end{align}
$\mathcal{D}_i>1$ means that the shift in bin $i$ is larger than the precision targeted by a stacked analysis. Note that this is not the same as the bias in $k_0$ when the lensed data are analysed without lensing. We compute that bias separately below.

Once lensing is included, the apparent turnover can no longer be treated as the same redshift-independent quantity across redshift bins. Stacking redshift bins then requires an accurate model of the lensing contribution.
\end{itemize}

For the fiducial turnover scale of the matter power spectrum, we use $k_0=0.0163\,h/$Mpc.

\subsection{Euclid-like survey}

We choose 4 bins of width $\Delta z = 0.225$, spanning $0.9 \leq z \leq 1.8$, with sky area 15,000~deg$^2$. For the standard case, we recover the turnover as
\begin{align}\label{snole}
   k_0 = 0.01645 \pm 0.00149\,h\,\mathrm{Mpc}^{-1} \qquad \mbox{(lensing neglected)}.
\end{align}
This result follows from combining the 4 redshift bins in the no-lensing case.
This precision is lower than the $\sigma(k_0)=0.00086\,h\,{\rm Mpc}^{-1}$ that we found in~\cite{Dube:2025hqf} for a similar survey. We checked this by starting from the present analysis and changing one choice at a time to that of~\cite{Dube:2025hqf}. The difference comes mainly from two choices. Firstly, the monopole variance used in~\cite{Dube:2025hqf} is a factor 2 smaller than the standard Gaussian result \autoref{eq:cov}, which we use here; this reduces $\sigma(k_0)$ by a factor $\approx 0.67$. Secondly, the redshift bins in~\cite{Dube:2025hqf} extended to $z=2.1$, which gives 39\% more volume and reduces $\sigma(k_0)$ by a factor $\approx 0.84$. Fitting one set of $(k_0,\alpha,\beta)$ jointly to all bins, as in~\cite{Dube:2025hqf}, instead of fitting each bin separately, gives a further reduction of $\approx 10\%$. With all these choices together we recover $\sigma(k_0)=0.00077\,h\,{\rm Mpc}^{-1}$. The survey specifications and the growth function change $\sigma(k_0)$ by less than 3\%.

\begin{figure}[!htbp]
  \centering
  \includegraphics[width=0.5\linewidth]%
    {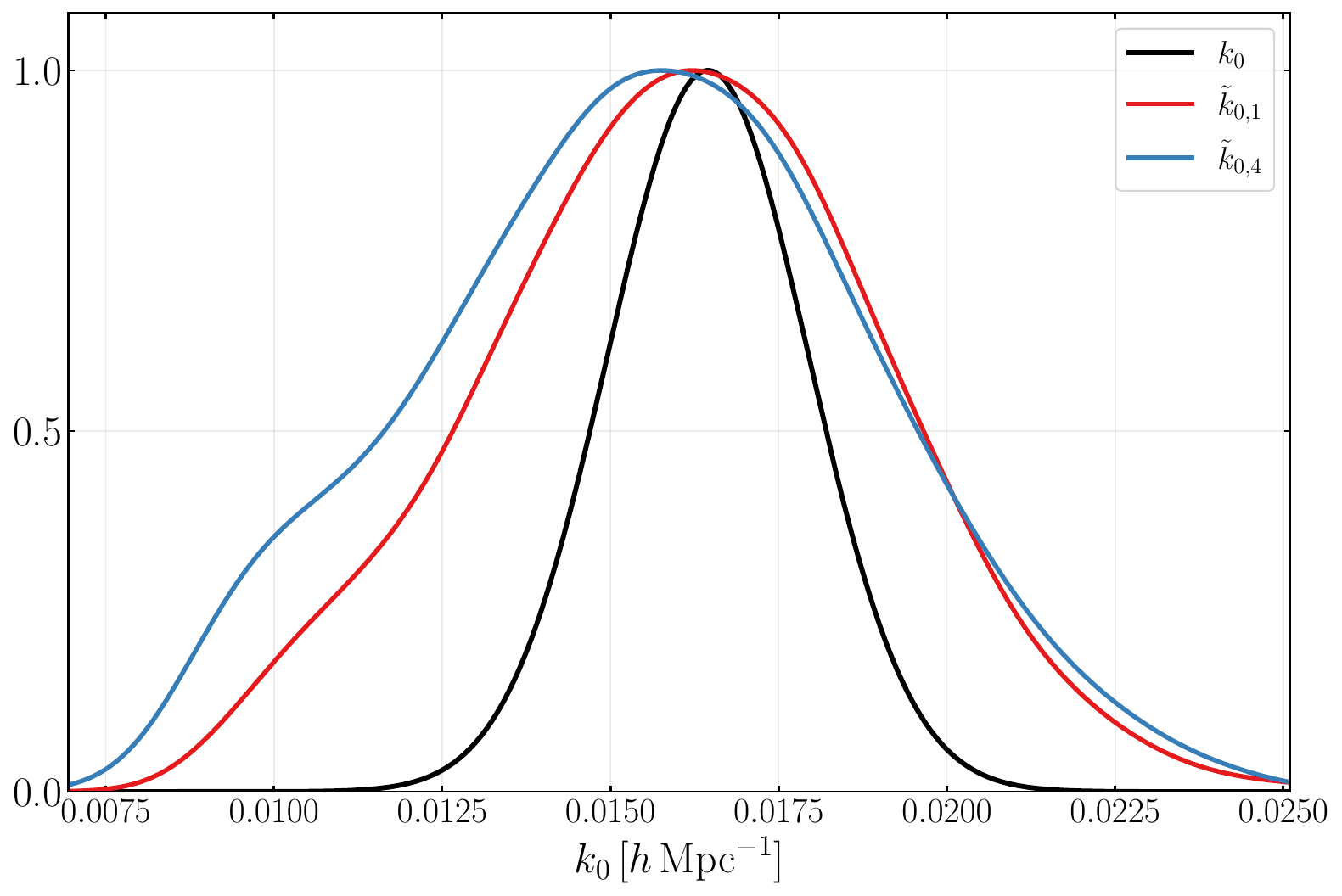}
  \caption{
    Marginal posteriors of the turnover scale for the Euclid-like survey. The black curve is the standard (no-lensing) posterior. Coloured curves are the per-bin posteriors of $\tilde k_{0,i}$, when lensing is included, shown at two representative redshifts, $z_1 = 1.01$ and $z_4=1.69$.
  }
  \label{fig:k0_post_eu}
\end{figure}

When lensing is included, the turnover scale changes with redshift. The results are summarised in \autoref{tab:lensing_bias_euclid} and \autoref{fig:k0_post_eu}.

\begin{table}[!htbp]
\centering
\begin{tabular}{ccccc}
\toprule
 bin & recovered turnover & error & fractional shift in turnover & shift in units of $\sigma(k_0)$ \\
 $z_i$ & $\tilde{k}_{0,i}$ & $\sigma(\tilde{k}_{0,i})$ & $[k_0 - \tilde{k}_{0,i}]/k_0$ & $\mathcal{D}_i$ \\
\midrule
        1.01 & 0.01623 & 0.00295 & 0.014 & 0.151 \\
        1.24 & 0.01633 & 0.00300 & 0.007 & 0.081 \\
        1.46 & 0.01622 & 0.00316 & 0.014 & 0.152 \\
        1.69 & 0.01601 & 0.00338 & 0.027 & 0.297 \\
\bottomrule
\end{tabular}
\caption{For a Euclid-like survey, the recovered per-bin turnover, when including lensing, and its shift in units of the no-lensing precision, \autoref{eq:D}. The recovered standard values of $k_0, \sigma(k_0)$ are given in \autoref{snole}.
}
\label{tab:lensing_bias_euclid}
\end{table}

\subsection{MegaMapper-like survey}

\begin{figure}[!htbp]
  \centering
  \includegraphics[width=0.5\linewidth]%
    {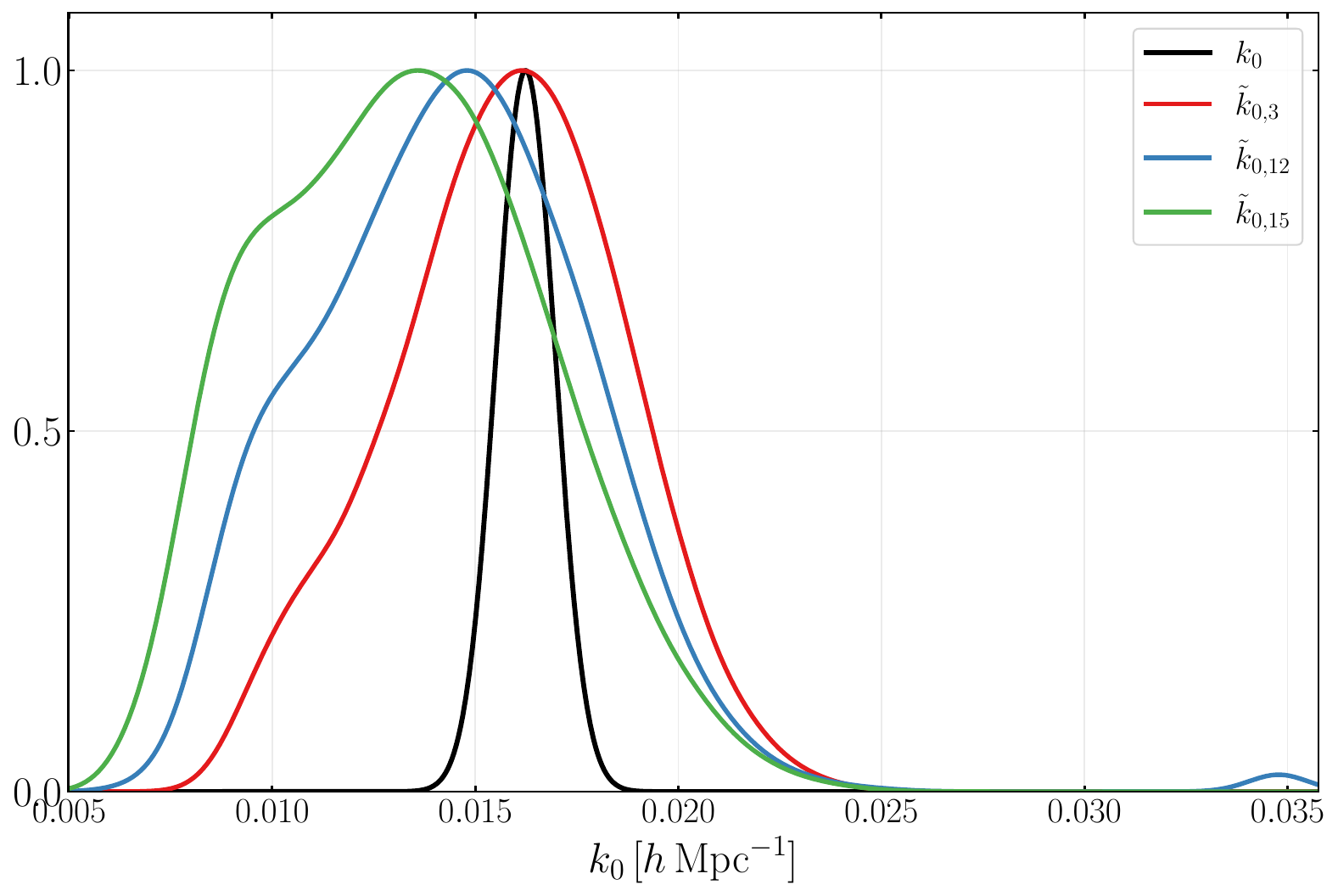}
  \caption{
    Marginal posteriors of the turnover scale for the MegaMapper-like survey. The black curve is the standard (no-lensing) posterior. Coloured curves are the per-bin posteriors $\tilde k_{0,i}$ when lensing is included, shown at three representative redshifts, $z_3 = 2.35,z_{12}= 3.25$ and $z_{15}=3.55$.
  }
  \label{fig:k0_post_mm}
\end{figure}

Here we use the redshift range $2.1 \leq z \leq 3.6$ and sky area 20,000~deg$^2$. Although the turnover persists until $z\approx 3.7$, it cannot be reliably recovered for $z>3.6$ -- since the power spectrum monopole becomes too flat at $k<\tilde k_0$ (see \autoref{fig4}). Given the strong evolution in $k<\tilde k_0(z)$, we choose 15 thinner bins, of width $\Delta z = 0.1$. In the standard case, we find
\begin{align}\label{snolm}
    k_0 = 0.01624 \pm 0.00073\,h\,\mathrm{Mpc}^{-1} \qquad \mbox{(lensing neglected)}.
\end{align}
When lensing is included, the per-bin recovered turnover is given in \autoref{tab:lensing_bias_megamapper} and \autoref{fig:k0_post_mm}.

\begin{table}[!htbp]
\centering
\begin{tabular}{ccccc}
\toprule
 bin & recovered turnover & error & fractional shift in turnover & shift in units of $\sigma(k_0)$ \\
 $z_i$ & $\tilde{k}_{0,i}$ & $\sigma(\tilde{k}_{0,i})$ & $[k_0 - \tilde{k}_{0,i}]/k_0$ & $\mathcal{D}_i$ \\
\midrule
        2.15 & 0.01611 & 0.00285 & 0.008 & 0.176 \\
        2.25 & 0.01611 & 0.00287 & 0.008 & 0.178 \\
        2.35 & 0.01608 & 0.00286 & 0.010 & 0.225 \\
        2.45 & 0.01603 & 0.00290 & 0.013 & 0.282 \\
        2.55 & 0.01599 & 0.00291 & 0.016 & 0.347 \\
        2.65 & 0.01592 & 0.00299 & 0.019 & 0.432 \\
        2.75 & 0.01583 & 0.00301 & 0.025 & 0.566 \\
        2.85 & 0.01570 & 0.00307 & 0.033 & 0.744 \\
        2.95 & 0.01541 & 0.00314 & 0.051 & 1.133 \\
        3.05 & 0.01538 & 0.00320 & 0.053 & 1.175 \\
        3.15 & 0.01510 & 0.00330 & 0.070 & 1.568 \\
        3.25 & 0.01480 & 0.00343 & 0.089 & 1.979 \\
        3.35 & 0.01450 & 0.00350 & 0.107 & 2.382 \\
        3.45 & 0.01395 & 0.00356 & 0.141 & 3.139 \\
        3.55 & 0.01367 & 0.00364 & 0.158 & 3.517 \\
\bottomrule
\end{tabular}
\caption{For a MegaMapper-like survey, the recovered per-bin turnover, when including lensing, and its shift in units of the no-lensing precision, \autoref{eq:D}. The recovered standard values of $k_0, \sigma(k_0)$ are given in \autoref{snolm}.}
\label{tab:lensing_bias_megamapper}
\end{table}

\begin{figure}[H]
  \centering
  \includegraphics[width=0.7\linewidth]%
    {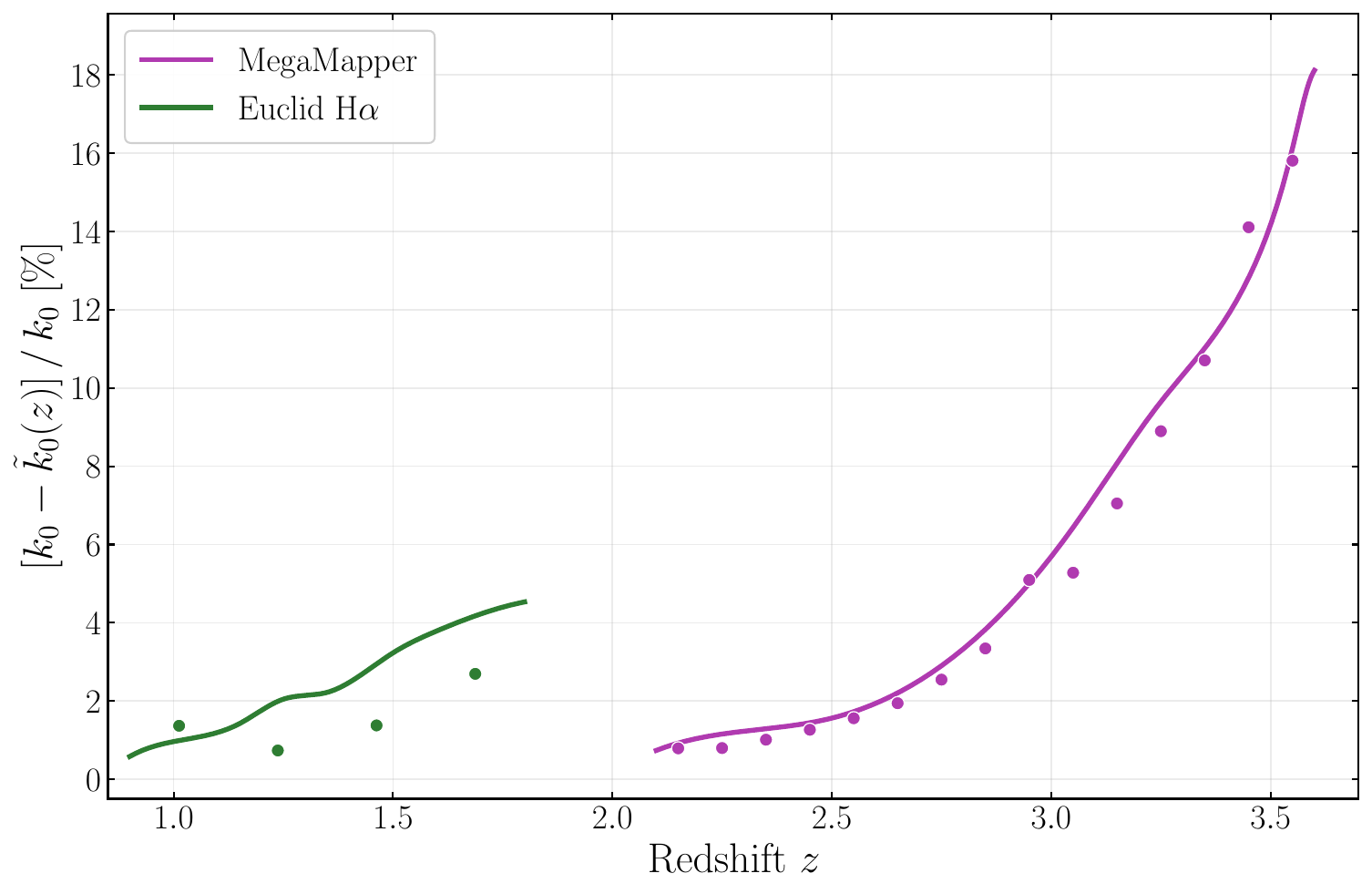}
  \caption{
    {Theoretical lensing-induced turnover shift (solid curves) compared with the MCMC recovery (dots), for the Euclid-like H$\alpha$ (green) and MegaMapper-like (magenta) surveys. Solid curves are smooth polynomial fits to the theoretical shift $[k_0 - \tilde k_0(z)]/k_0$ (in \%). Dots are the per-bin shifts $[k_0 - \tilde k_{0,i}]/k_0$ recovered from the MCMC, as shown in \autoref{tab:lensing_bias_euclid} and \autoref{tab:lensing_bias_megamapper}. }
  }
  \label{fig:k0_shift_mcmc}
\end{figure}

In \autoref{fig:k0_shift_mcmc}, we compare the shift on the turnover induced by lensing, which is recovered from MCMC simulations, with the fiducial shift from theory, which is taken from \autoref{fig3} and \autoref{fig4}. The recovered points track the theoretical curves across the full redshift range, confirming that the per-bin fits capture the expected lensing shift. Since the mock data are noiseless, the scatter of the points about the curves comes from the finite MCMC sampling. For a single chain, the maximum of the per-bin posterior varies by $\sim2\times10^{-4}\,h/$Mpc between runs, i.e.\ $\sim0.3$ in $\mathcal{D}_i$. To reduce this, the values in \autoref{tab:lensing_bias_euclid} and \autoref{tab:lensing_bias_megamapper} are obtained from 8 independent chains per bin. The noise-free theory curve gives $\mathcal{D}>1$ for $z\geq2.925$.

\subsection{Robustness of the results}

\noindent{\em Stacking lensed data without lensing}\\

\noindent  $\mathcal{D}_i$ compares the shift in a single bin with the precision of the stacked analysis. A different question is how biased $k_0$ would be if the lensed data were analysed as in the standard case, i.e.\ by stacking all bins as if the turnover were independent of redshift. We answer this by applying \autoref{eq:stack} to the lensed per-bin results, and comparing with the standard result, as a function of the maximum redshift of the stacked bins (\autoref{tab:misspecified}). For Euclid-like, the bias is $-0.18\sigma$. For MegaMapper-like, it is $-0.24\sigma$ for $z\le2.9$, and reaches $-0.93\sigma$ when all bins up to $z=3.6$ are included. The lensed per-bin values are consistent with a common value ($\chi^2=0.8$ for 14 degrees of freedom), since the error in each bin is $\sim4$ times larger than the shift across the bins. The redshift dependence would therefore not be noticed in a stacked analysis.\\

\begin{table}[!htbp]
\centering
\begin{tabular}{cccccc}
\toprule
survey & $z_{\max}$ & bins & $k_0$ (no lensing) & $k_0$ (lensing, stacked) & bias \\
\midrule
Euclid-like & 1.8 & 4 & $0.01647\pm0.00149$ & $0.01621\pm0.00155$ & $-0.18\sigma$ \\
\midrule
MegaMapper-like & 2.5 & 4 & $0.01621\pm0.00141$ & $0.01608\pm0.00143$ & $-0.09\sigma$ \\
 & 2.9 & 8 & $0.01621\pm0.00100$ & $0.01598\pm0.00104$ & $-0.24\sigma$ \\
 & 3.2 & 11 & $0.01620\pm0.00086$ & $0.01582\pm0.00090$ & $-0.45\sigma$ \\
 & 3.6 & 15 & $0.01619\pm0.00075$ & $0.01549\pm0.00080$ & $-0.93\sigma$ \\
\bottomrule
\end{tabular}
\caption{Turnover recovered by stacking the redshift bins up to $z_{\max}$, using \autoref{eq:stack}, for the standard and lensed data. The bias is the difference in units of the standard $\sigma(k_0)$. The standard values are from the same MCMC chains as the lensed values, and agree with \autoref{snole} and \autoref{snolm} within the sampling noise.}
\label{tab:misspecified}
\end{table}

\noindent{\em Fitting window and binning}\\

\noindent
For the MegaMapper-like survey, we computed $P^{(0)}_{\rm obs}$ over $0.0015\le k\le0.06\,h/$Mpc, and repeated the search for the turnover in different windows. The disappearance redshift does not depend on $k_{\rm max}$ in the range $0.030$--$0.045\,h/$Mpc. Lowering $k_{\rm min}$ makes the turnover disappear earlier: $z=3.65$ for $k_{\rm min}=0.005$, $3.575$ for $0.004$ and $3.5$ for $k_{\rm min}=k_{\rm f}$, compared with 3.675 for $k_{\rm min}=0.006\,h/$Mpc (on a grid with $\Delta z=0.025$). This is because of the low-$k$ upturn in \autoref{fig4}. The disappearance redshift also depends on the 2\% prominence that we require: without this requirement it is $z=3.78$, while for 5\% it is 3.58.

We also repeated the MCMC fits with $k_{\rm min}=k_{\rm f}$, with $k_{\rm max}=0.030$ and 0.045, and with $\Delta k=k_{\rm f}$ and $3k_{\rm f}$. Without lensing, the recovered $k_0$ ranges from 0.01539 to 0.01686$\,h/$Mpc, compared with the turnover of the standard monopole, $0.01639\,h/$Mpc. These values differ by up to $\sim2\sigma(k_0)$. This is a systematic of the phenomenological parabola, \autoref{eq:parabola}: the recovered $k_0$ depends on how much of the spectrum on each side of the peak is included in the fit. The precision $\sigma(k_0)$ ranges from 0.00059 to 0.00111$\,h/$Mpc, following the number of $k$-values per bin (4--10). Since the standard and lensed data are fitted in the same window, the shift is more stable than $k_0$ itself: $\mathcal{D}_i$ first exceeds 1 at $z_i=2.95$--3.35, and, for the settings with at least 6 $k$-values per bin, the stacked bias for $z\le3.6$ is between $-1.2\sigma$ and $-0.4\sigma$. Above $z\approx3.4$ the per-bin fits are not robust to these choices, which supports our choice of $z\le3.6$ for the MegaMapper-like analysis.\\

\noindent{\em Magnification bias}\\

\noindent Our results depend on $s(z)$ through the factor $5s-2$. In \autoref{tab:s_sensitivity} we rescale $s(z)$ by $\pm10\%$ and $\pm20\%$, change the magnitude limit to $m_c=24.0$ and 25.0 (which changes $s$, $b$ and $n_g$ consistently), and change the interpolation of the luminosity function between redshifts. A 10\% change in $s$ moves the redshift at which $\mathcal{D}>1$ by $\Delta z\approx0.13$, and the disappearance redshift by $\Delta z\approx0.2$--0.4. For a 20\% lower $s$, or $m_c=25.0$, the turnover does not disappear below $z=4.5$. In all cases, $\mathcal{D}$ exceeds 1 in the range $2.7\lesssim z\lesssim3.3$.\\

\begin{table}[!htbp]
\centering
\begin{tabular}{lcccc}
\toprule
$s(z)$ model & $s(3)$ & $\tilde k_0(3)\ [h/{\rm Mpc}]$ & $\mathcal{D}>1$ for $z\geq$ & disappearance $z$ \\
\midrule
fiducial ($m_c=24.5$) & 1.02 & 0.01548 & 2.925 & 3.675\\
$0.8\,s$ & 0.81 & 0.01600 & 3.300 & $>4.5$\\
$0.9\,s$ & 0.92 & 0.01577 & 3.075 & 4.075\\
$1.1\,s$ & 1.12 & 0.01516 & 2.800 & 3.475\\
$1.2\,s$ & 1.22 & 0.01477 & 2.675 & 3.300\\
$m_c=24.0$ & 1.33 & 0.01484 & 2.725 & 3.275\\
$m_c=25.0$ & 0.81 & 0.01585 & 3.125 & $>4.5$\\
linear interpolation & 1.02 & 0.01548 & 2.900 & 3.700\\
\bottomrule
\end{tabular}
\caption{Sensitivity of the MegaMapper-like results to the magnification bias, from the theoretical turnover $\tilde k_0(z)$ (redshift step 0.025). $\mathcal{D}$ uses $\sigma(k_0)$ from \autoref{snolm}.}
\label{tab:s_sensitivity}
\end{table}

\noindent{\em Other relativistic effects}\\

\begin{figure}[!htbp]
  \centering
  \includegraphics[width=\linewidth]{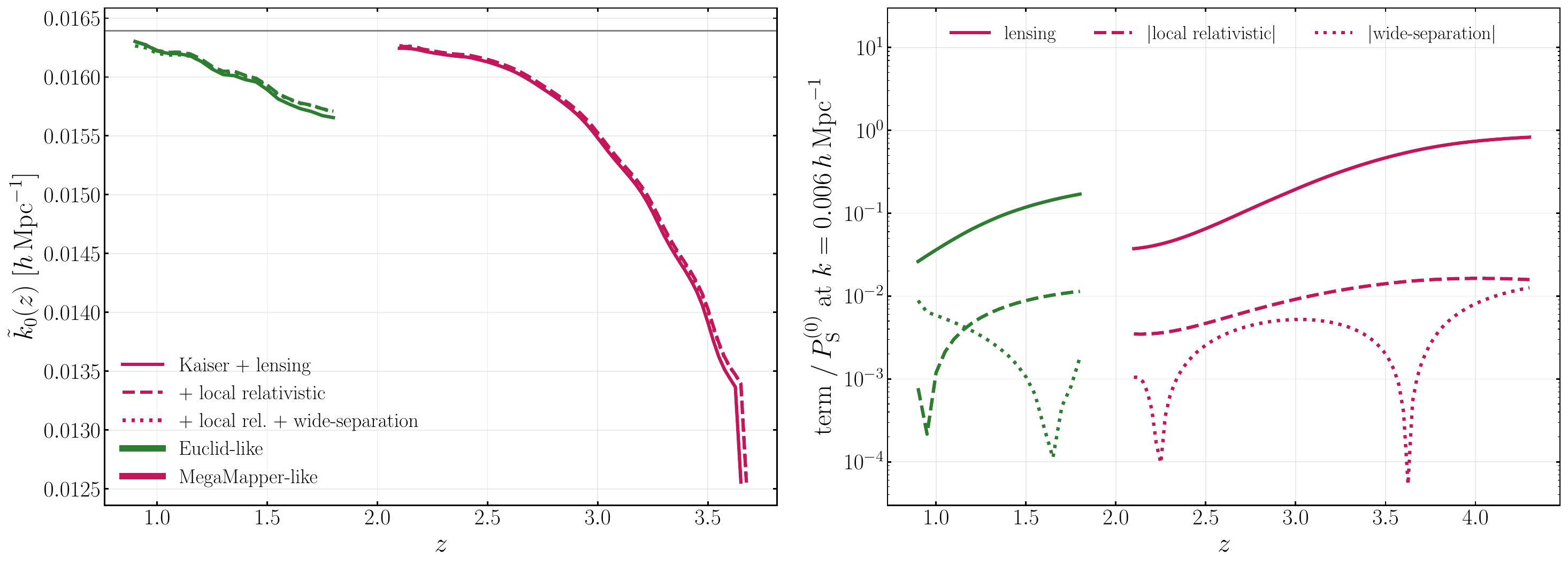}
  \caption{{\em Left:} Theoretical turnover $\tilde k_0(z)$ with lensing only (solid), with local relativistic terms added (dashed), and with local relativistic and wide-separation terms added (dotted), for Euclid-like (green) and MegaMapper-like (magenta) surveys. The grey line is the fiducial $k_0$. {\em Right:} Size of each correction relative to $P^{(0)}_{\rm S}$ at $k=0.006\,h/$Mpc.}
  \label{fig:relativistic}
\end{figure}

\noindent
Lensing is the dominant relativistic correction for these surveys, but the local relativistic (Doppler and potential) terms and the wide-separation (wide-angle and unequal-time) terms are not always negligible~\cite{Addis:2025rre}. We added these terms to $P^{(0)}_{\rm obs}$, using \texttt{CosmoWAP}, and recomputed the turnover (\autoref{fig:relativistic}). At $k=0.006\,h/$Mpc, the local terms reduce the monopole by 0.5--1.5\%, compared with the 7--56\% increase from lensing for $2.5\le z\le3.65$. They therefore partly cancel the lensing effect, but only at the level of 3--8\%. For MegaMapper-like, the theoretical turnover gives $\mathcal{D}>1$ for $z\geq2.95$, and disappears at $z=3.70$, compared with 2.925 and 3.675 for lensing only. For Euclid-like, the maximum of $\mathcal{D}$ from the theoretical turnover changes from 0.50 to 0.46.

\subsection{DESI Quasars}

Finally, we consider an existing catalogue that reaches high redshift. The Dark Energy Spectroscopic Instrument (DESI) Data Release 1 (DR1) includes a catalogue of more than 1.2 million spectroscopic quasars, in the redshift range $0.8 \le z \le 3.5$ and across 7,200~deg$^2$ \cite{deBelsunce:2025qku}.

\begin{figure}[H]
  \centering
  \includegraphics[width=0.7\linewidth]%
    {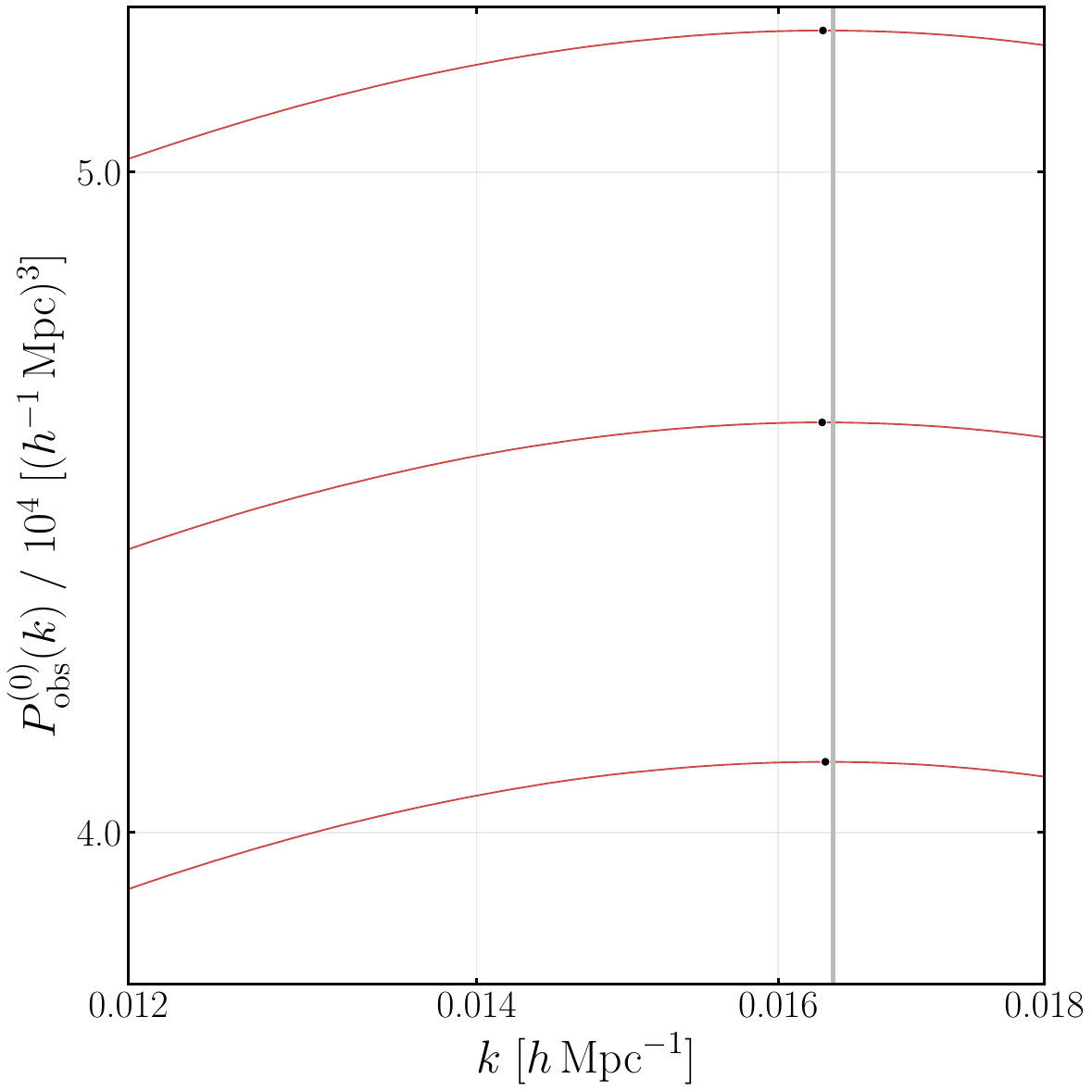}
  \caption{
    {Lensing-corrected power spectrum monopole $P^{(0)}_{\rm obs}(k)$ for the DESI DR1 quasar sample at three effective redshifts, $z = 1.55,\ 2.05$ and $2.55$. Black dots mark the turnover, shifted by lensing, and the grey vertical line is the fiducial turnover scale $k_0$.}
}
  \label{fig:turnover_desi}
\end{figure}

With such high redshifts, the lensing effects on the monopole should be investigated. Here we simply illustrate the theoretical shift, based on the clustering bias and magnification bias values given at 3 effective redshifts in \cite{deBelsunce:2025qku}. Then we compute the theoretical power spectrum at the effective redshifts. The impact of lensing on the monopole is shown in \autoref{fig:turnover_desi}. The magnification bias of the quasars is small, $0.10\le s\le0.24$, so that $5s-2<0$ and the lensing contribution is only 1--2\% of the monopole at $k=0.006\,h/$Mpc. The turnover is shifted by less than $10^{-4}\,h/$Mpc. Given the estimates of $s(z)$ in \cite{deBelsunce:2025qku}, a tentative conclusion is that lensing has a negligible effect. This is worth further investigation, as the catalogue is developed.

\section{Conclusion}
\label{sec:conclusion}

The matter power spectrum peaks at a particular turnover scale $k_0$, which encodes the effect of the radiation-to-matter transition on the evolution of modes of the matter density contrast. The turnover scale contains information about the matter-radiation equality scale and is effectively a standard ruler that can constrain cosmological parameters, serving as a complement to observations such as BAO.

The standard (Kaiser) galaxy power spectrum monopole, as a tracer of the matter power spectrum, thus carries the same information, since the turnover scale is independent of redshift evolution. Thus, the turnover scale can be constrained more precisely by stacking more redshift bins in any particular observation.

As examples, we investigate two mock surveys based on Euclid H$\alpha$ and MegaMapper. If we ignore lensing, the Euclid-like survey delivers the constraint $\sigma(k_0) = 0.00149\,h\,\mathrm{Mpc}^{-1}$, when four redshift bins (of width $\Delta z = 0.225$) are stacked. The Megamapper-like surveys greater constraining power, with $\sigma(k_0) = 0.00073\,h\,\mathrm{Mpc}^{-1}$, {when 15 redshift bins (of width $\Delta z = 0.1$) are stacked}  within the redshift range $2.1 \leq z \leq 3.6$.

However, these very tight constraints rely on the Kaiser model of the galaxy power spectrum -- which is not the observed galaxy power spectrum due to relativistic effects. In this analysis, we showed that lensing magnification has a non-negligible effect on the galaxy power spectrum and its turnover. Hence, the turnover scale is biased away from the matter power spectrum turnover $k_0$. Lensing magnification shifts the peak position of the observed galaxy power spectrum in different ways at different redshifts, so that the turnover scale and the corresponding bias are not fixed but redshift-dependent.

The measured turnover scale is redshift-dependent, and the observed galaxy power spectrum monopole has a distorted shape around its peaks. Consequently, the apparent turnover can no longer be treated as the same redshift-independent quantity across redshift bins, and stacking the bins requires an accurate lensing model. Thus, we compute the corrected turnover scale and the associated shift at each redshift bin. Assuming $s(z)$ is known exactly, we compute these shifts, in units of the stacked no-lensing precision, for both of the mock surveys. We find $\mathcal{D}\sim 0.3$ at $z\sim1.69$ (Euclid-like) and $\sim 3.5$ at $z=3.55$ (MegaMapper-like). If the MegaMapper-like bins up to $z=3.6$ are stacked without lensing, the bias on $k_0$ is $-0.93\sigma$. While we may neglect the shifts for the Euclid-like survey, we definitely cannot ignore them for the MegaMapper-like survey. We also computed the theoretical shift of the turnover for a DESI DR1 quasar sample, using DR1 estimates, and found no significant shift.

The redshifts $z\approx2.9$ (above which $\mathcal{D}>1$) and $z\approx3.7$ (above which the turnover disappears) for MegaMapper-like depend on our assumptions. They are sensitive to the magnification bias: a 20\% change in $s(z)$ moves them to $2.7\lesssim z\lesssim3.3$ and from $z\approx3.3$ to beyond $z=4.5$, respectively. The disappearance redshift also depends on the fitting window and on how we define a detectable turnover, which gives $3.5\lesssim z\lesssim3.8$. These redshifts are, however, insensitive to the local relativistic and wide-separation corrections that we neglected, which change them by $\Delta z\le0.025$.

In~\cite{Dube:2025hqf} we showed that adding the equilateral bispectrum monopole improves the precision on $k_0$, assuming that it shares the redshift-independent turnover of the power spectrum. Lensing magnification also contributes to the observed bispectrum, so that its turnover will also be shifted. This gain is therefore reliable only where lensing is negligible, as for Euclid-like redshifts, and needs to be revisited for MegaMapper-like surveys.

Our forecasts use noiseless mock data and a Gaussian covariance without the survey window. On the scales of the turnover, the window changes both the measured monopole and its covariance, and correlates neighbouring $k$-bins. Since the lensing signal is largest at the smallest $k$, the window will interact with the lensing shift. A realistic measurement will also need to account for super-sample covariance and for observational systematics, such as imaging systematics and spatially varying selection. We leave these to future work.

In summary, either we lose information on the constraints on the cosmological parameters, which are obtained through the turnover detection of the observed galaxy power spectrum when the lensing effect is correctly accounted for, or we need to do proper modeling of magnification bias.

\vfill
\acknowledgments
YD, RM and BRD were supported by the South African Radio Astronomy Observatory and the National Research Foundation (grant no.\ 75415).
We thank Chris Addis for useful discussions on  \texttt{CosmoWAP} implementation.

\clearpage
\bibliographystyle{JHEP}
\bibliography{References}

\end{document}